\documentclass[fleqn,usenatbib]{mnras}

\usepackage{newtxtext,newtxmath}

\usepackage[T1]{fontenc}

\usepackage{graphicx}	
\usepackage{amsmath}	
\usepackage[utf8]{inputenc}  
\usepackage[T1]{fontenc}
\usepackage{mathrsfs}
\usepackage{multirow,enumerate}
\usepackage{comment,hyperref}
\usepackage{color}
\usepackage{acronym}
\usepackage{xspace}
\usepackage[normalem]{ulem}
\usepackage{mathtools}
\usepackage{subfigure}
\usepackage{makecell}

\title{Robust parameter estimation from pulsar timing data}

\author[A. Samajdar et al.]{
A.~Samajdar,$^{1,2}$\thanks{E-mail:anuradha.samajdar@uni-potsdam.de}
G.~Shaifullah,$^{1}$
A.~Sesana,$^{1,3,4}$
J.~Antoniadis,$^{5,6}$
M.~Burgay,$^{7}$
\newauthor
D.~J.~Champion,$^{5}$
S.~Chen,$^{8}$
M.~Kramer,$^{5,9}$
J.~W.~McKee,$^{10}$
M.~B.~Mickaliger,$^{9}$
\newauthor
E.~Van der Wateren$^{11,12}$
\\
$^{1}$Department of Physics, University of Milano -- Bicocca, 
Piazza della Scienza 3, 20126 Milano, Italy\\
$^{2}$Institut f\"{u}r Physik und Astronomie, Universit\"{a}t Potsdam, Haus 28, 
Karl-Liebknecht-Str. 24/25, 14476, Potsdam, Germany\\
$^{3}$INFN, Sezione di Milano-Bicocca, Piazza della Scienza 3, I-20126 Milano, Italy\\
$^{4}$INAF - Osservatorio Astronomico di Brera, via Brera 20, 20121 Milano, Italy\\
$^{5}$Max-Planck-Institut f\"{u}r Radioastronomie, Auf dem H\"{u}gel 69, 53121 Bonn, Germany\\
$^{6}$Institute of Astrophysics, Foundation for Research \& Technology -- Hellas (FORTH), GR-70013 Heraklion, Greece\\
$^{7}$INAF - Osservatorio Astronomico di Cagliari, via della Scienza 5, 09047 Selargius (Cagliari), Italy\\
$^{8}$Kavli Institute for Astronomy and Astrophysics, Peking University, 100871 Beijing, P. R. China\\
$^{9}$Jodrell Bank Centre for Astrophysics, School of Physics and Astronomy, The University of Manchester, Manchester M13 9PL, UK\\
$^{10}$Canadian Institute for Theoretical Astrophysics, University of Toronto, 60 Saint George Street, Toronto, ON M5S 3H8, Canada\\
$^{11}$ASTRON, the Netherlands Institute for Radio Astronomy, Postbus 2, 7990 AA Dwingeloo, The Netherlands\\
$^{12}$Department of Astrophysics/IMAPP, Radboud University, PO Box 9010, 6500 GL Nijmegen, The Netherlands
}

\date{Accepted XXX. Received YYY; in original form ZZZ}

\pubyear{2022}

\begin{document}
\label{firstpage}
\pagerange{\pageref{firstpage}--\pageref{lastpage}}
\maketitle

\begin{abstract}
Recently, global pulsar timing arrays have released results from searching for a nano-Hertz gravitational wave background signal. Although there has not been any definite evidence of the presence of such a signal in residuals of 
pulsar timing data yet, with more and improved data in future, a statistically significant detection is expected to be made. 
Stochastic algorithms are used to sample a very large parameter space to infer results from data. In this paper, we attempt to rule 
out effects arising from the stochasticity of the sampler in the inference process. 
We compare different configurations of nested samplers and the more commonly used markov chain monte carlo method to sample the 
pulsar timing array parameter space and account for times taken by the different samplers on same data. 
Although we obtain consistent results on parameters from different sampling algorithms, we propose 
two different samplers for robustness checks on data in the future to account for cross-checks between sampling methods as well as 
realistic run-times.
\end{abstract}




\section{Introduction}
\label{sec:intro}
Pulsars Timing Arrays (PTAs) (\cite{Detweiler:1979wn, Hellings:1983fr}) aim to detect the stochastic Gravitational Wave Background (GWB). 
A GWB signal is likely 
created by the superposition of gravitational waves emitted by Super Massive Black Hole Binaries (SMBHBs) (\cite{Rosado:2015epa}), but there could be other sources like a relic from inflation (\cite{Grishchuk_2005}) or cosmic strings (\cite{VILENKIN198147, vilenkin94}). While increasingly constraining upper limits have been placed on the amplitude of the GWB there has been no detection of this signature yet. However all operational PTAs are currently detecting a common but spatially
uncorrelated red noise process (\cite{Ferdman:2010xq, Manchester:2012za, Hobbs:2009yy, Jenet:2009hk}). This might indicate that the GWB signal will be 
detected with statistical significance in the near future. In this paper, we look at consistencies 
between a variety of stochastic samplers used to sample a large parameter space, where the latter is of paramount importance to inferring properties of the GWB signal. 

The size of pulsar timing models necessitates the use of a hybrid frequentist and Bayesian analysis, where the pulsar timing model parameters are first obtained using iterative least square fitting with tools such as \texttt{TEMPO2} (\cite{Hobbs:2006cd, Edwards:2006zg, Hobbs:2009yn}) or 
PINT (\cite{Luo:2020ksx}) to obtain a set of timing residuals. These timing residuals are then modelled to remove excess delays due to red noise processes as well as fluctuations from the variations in the ionised interstellar medium codified as dispersion-measure models using Bayesian analysis, while analytically marginalising over the timing model parameters. This is typically called single pulsar noise analysis (SPNA). Even with this simplification, the estimation of the red noise and dispersion-measure model parameters remains computationally expensive. Further, in the final stage, when searching for the GWB, all pulsar models must be simultaneously fitted for, along with a model for the correlated signal from the GWB as well as any other correlated or uncorrelated common noise processes. Even when the search is optimised for the smallest number of pulsars, this can lead to final dimensions of the order of hundreds of parameters. Further, for the individual pulsar models as well as the final GWB search, it is desirable to carry out model selection (\cite{jeffrey,Raftery}) -- a method which is particularly well suited for Bayesian analysis.

This mixed approach implies inherent uncertainties in the comparision of the algorithms themselves as well as any difference in the obtained results. We attempt to address this issue by adapting the most commonly used PTA analysis package, enterprise, to utlise a number of nested sampling algorithms. We perform single pulsar noise analyses for a set of six pulsars, first utlised for the recent limits presented by the  European Pulsar timing Array (EPTA) (\cite{Chen:2021rqp}). Using the most performant nested sampling algorithm as determined from the SPNA analysis, we then search for the GWB using both the pulsar dataset as well as the second International Pulsar Timing Array (IPTA) second mock data challenge (MDC) (\cite{Hazboun:2018wpv}).

In Sec.~\ref{sec:data}, we briefly summarise the data we have used for our inference process. Section~\ref{sec:method} serves 
as an introduction to Bayesian inference with focus on noise models used in this paper and pulsar timing data in general. Some technical details 
to algorithms we use are also included. We give a summary of our results in Sec.~\ref{sec:results} and conclude in Sec.~\ref{sec:conclusion}.

\section{Data}
\label{sec:data}
\begin{figure}
    \centering
    \includegraphics[scale = 0.5]{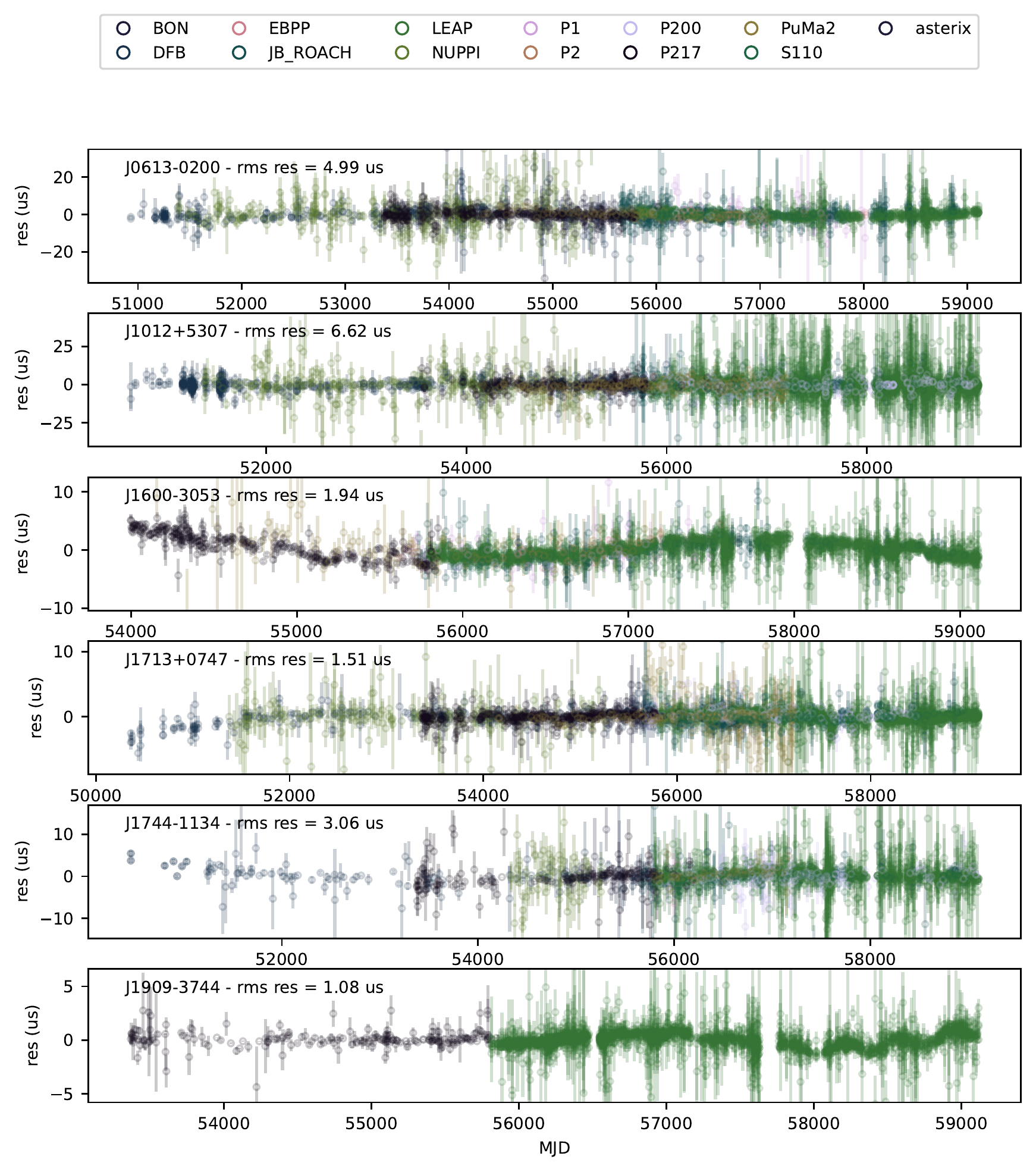}
    \caption{Timing residuals for the 6 pulsars from \protect\cite{Chen:2021rqp}. Colors denote the different data recording systems (or backends) used and the labels are described in the text.}
    \label{fig:residuals}
\end{figure}

We have used data recently utilised by the EPTA collaboration (\cite{Chen:2021rqp}), and focussed 
on six pulsars -- PSRs~J0613-0200, J1012+5307, J1600-3053, J1713+0747, J1744-1134 and J1909-3744. The times of arrival (TOAs) of these pulsars are fitted using the \texttt{TEMPO2} software to pulsar timing models describing the pulsars astrometric, intrinsic and environmental properties, along with simple polynomial models for the variations of ionised interstellar medium (IISM) along the line of sight to these pulsars. The resulting `timing residuals' are shown in Fig.~\ref{fig:residuals} where we highlight the large number of observing systems used for each pulsar dataset by different colours. We refer the interested reader to (\cite{Chen:2021rqp}) and forthcoming EPTA publications for more details on the individual observing systems but list the names here. the abbreviations correspond to the Pulsar Machine (P1, P2 and PuMaI/II) instruments at the Westerbork Synthesis Radio Telescope (WSRT), the Reconfigurable Open Architecture Computing Hardware (ROACH) and the Digital Filter Bank (DFB) based devices at the Jodrell Bank Observatory (JBO), the Berkeley-Orleans-Nan\c{c}ay (BON) and the Nancay Ultimate Pulsar Processign Instrument (NUPPI) at the Nan\c{c}ay Radio Observatory, and the PSRIX instrument (labelled P217, P200, S110 and asterix) at the Effelsberg radio telescope. 
The residuals of Fig.~\ref{fig:residuals} encode within them the signatures of contributions from pulsar specific low and high frequency processes as well as common astrophysical signals, such as perturbations due to Solar system bodies (\cite{Champion:2010zz,Caballero:2018lvc}), time-variable delays due to density fluctuations in the IISM along the line of sight to the pulsar or the spatially-correlated GWB.

In addition to real data, to test different samplers, we have also used simulated timing data, generated by the 
IPTA collaboration (\cite{Verbiest:2016vem}) and used in the second MDC (\cite{Hazboun:2018wpv}). 
From the MDC, we choose a dataset containing a GWB signal. The dataset consists of 33 pulsars, and 
in addition to the GWB, each individual pulsar is characterised by its own spin noise or red noise. 
The data also contains white noise characterising the observing telescopes. 
The simulated dataset spans a timeline of 15 years, and is observed at a central frequency of 1440 MHz. 
The times of arrival (TOAs) are uniformly distributed with obsevations taken every 30 days. 

The extraction of the GWB signal is a complicated process due to the need to transform radio pulsar observations 
into reference times at which a group of photons from each pulsar in the PTA arrive at Earth or Solar System Barycentre. 
While the observed data are the TOAs, the data analysis is done on timing residuals. For this the TOAs are first converted into residuals, 
obtained after subtracting the predicted timing model from the observed TOAs. If the predicted model fits the observations perfectly, the residuals 
will be identically 0. In addition to the presence of a GWB, additional non-gravitational-wave related noise sources may alter the 
TOAs, some of these noise models are described in Sec.~\ref{sec:method}.

\section{Bayesian analysis and parameter estimation setup}
\label{sec:method}
We perform Bayesian inference on the pulsar timing data and sample over parameters corresponding to noise models describing 
the variations in the residuals as described below. 
We sample over single pulsars (henceforth, SPNA analysis) as well as 
the full pulsar timing array (henceforth, PTA analysis) and use different samplers to test 
the consistency of the inferred noise models. 
Although nested sampling would appear to be naturally desirable for PTA analysis, analyses such as those from the Parkes PTA (PPTA) or NANOGRAV typically utilise the Parallel Tempering Markov Chain Monte Carlo (PTMCMC) (\cite{justin_ellis_2017_1037579}) method due to the lower computational cost. 
Since PTMCMC does provide direct access to the marginal likelihoods, 
methods such as the Savage-Dickey approximation and hypermodel sampling are employed for model comparison. Even though the EPTA and IPTA results have been presented in the literature which utilise efficient multi-ellipsoidal nested sampling algorithms such as \texttt{MultiNest} and \texttt{Polychord} for SPNA, the final search for the GWB still utilises PTMCMC or similar Markov Chain Monte Carlo (MCMC) based methods.
\subsection{Bayesian inference}
We provide a very brief summary of Bayesian inference to make our study self-contained. 
We point the reader to detailed resources like (\cite{sivia2006, gregory}) for further reading. 
Bayesian analysis estimates parameters from probability distribution functions (PDFs). The \emph{posterior} PDF is obtained 
by providing the initial \emph{prior} PDF and using that in combination with the \emph{likelihood}, containing information about the data. 
The Bayes' theorem can be written down as 
\begin{equation}
 P(\vec{\theta}|d) = \frac{P(d|\vec{\theta})P(\theta)}{P(d)}.
 \label{eq:bt}
\end{equation}
where $\vec{\theta}$ refers to a multidimensional parameter set, $d$ is the data and the notation $P(\vec{\theta}|d)$ 
refers to information on $\theta$ given $d$. Deatils of the likelihood calculation in case of 
analysis of pulsar timing data 
may be found in (\cite{NANOGrav:2015qfw}) and references therein. 
In addition to estimating parameters by using prior knowledge as well as knowledge from observed data, Bayesian analysis allows us to 
perform \emph{model selection}. With the data remaining the same, this means performing an analysis each time with a different model. 
In that case, Eqn.~\ref{eq:bt} may be rewritten as 
\begin{equation}
P(\mathcal{H}|d) = \frac{P(d|\mathcal{H})P(\mathcal{H})}{P(d)}.
\label{eq:bt_hyp}
\end{equation}
$\mathcal{H}$ represents a hypothesis and in case of pulsar analysis, $\mathcal{H}$ may be assuming that the timing data contains a GWB 
signal, $\mathcal{H}_\mathrm{GWB}$ or only a common red noise signal (CRN), $\mathcal{H}_\mathrm{CRN}$ but not a GWB. 
From Eqn.~\ref{eq:bt_hyp}, if we then compute the ratio of probabilities $P(\mathcal{H}_\mathrm{GWB})$ and $P(\mathcal{H}_\mathrm{CRN})$ 
we get a quantitative measure of which model is more preferred by the data.
\subsection{Noise models}
When analysing single pulsar data, we focus only on individual pulsars' intrinsic red noise (RN), the noise from 
disperson measure arising from the interstellar medium (DM), and white noise (WN), inherent to the 
observing telescopes. In addition, when sampling over the parameter space of a PTA analysis, we include 
a CRN. When the common process includes spatial correlations, we search for 
a common GWB. We briefly describe each of the noise processes below:
\subsubsection{White Noise}
The white noise itself can be divided into two parts: (i) a multiplicative factor of the estimated error bar on the observed TOAs, the $\mathrm{EFAC}$ and 
(ii) an additional noise adding in quadrature to the error bars, the $\mathrm{EQUAD}$. Both these components 
vary across the different observing telescopes even if they all observe the same pulsar. 
The total error on a TOA, $\sigma$ can be written as
\begin{equation}
 \sigma = \sqrt{(\sigma\mathrm{EFAC})^2 + \mathrm{EQUAD}^2}.
\end{equation}
$\mathrm{EFAC}$ represents possible uncertainty on the TOA error estimation during the cross-correlation of the pulsar profile with the stndard template (\cite{Taylor92}), and $\mathrm{EFAC}$ may arise from physical effects like pulsar jitter noise and give rise to additional scatter of the TOAs (\cite{equad}). 
\subsubsection{Red Noise}
Red noise is intrinsic to each pulsar, and also commonly referred to as spin-noise. 
This arises primarily as a result of irregularilities in pulsar-spin (\cite{RN, RN1}). 
The imprint on the pulsar residuals from the intrinsic noise is also a red process, like the GWB, and the power spectrum 
may be described as a powerlaw
\begin{equation}
 \phi_{\mathrm{RN}} = \frac{A_{\mathrm{RN}}^2}{12\pi^2} \left( \frac{1}{1 \ \mathrm{yr}} \right)^{-3} \frac{f^{-\gamma_{\mathrm{RN}}}}{T},
\end{equation}
where $A_{\mathrm{RN}}$ and $\gamma_{\mathrm{RN}}$ are the amplitude and spectral index of the red noise process respectively, and $T$ is the total timespan between latest and earliest TOA. 
\subsubsection{Dispersion Measure Noise}
As the pulses from a pulsar travel through the interstellar medium, the imprint of the interstellar medium is 
also encoded on the TOAs. Dispersion measure is time-varying and defined as the integrated column density of free electrons in
the pulsar's line of sight (\cite{You:2007fi}). 
Unlike intrinsisc red noise, this noise is frequency-dependent and follows a $\nu^{-2}$ dependence, $\nu$ being the radio frequency. This source maybe further described by an additional powerlaw spectrum of the form
\begin{equation}
 \phi_{\mathrm{DM}} = A_{\mathrm{DM}}^2 \left( \frac{1}{1 \ \mathrm{yr}} \right)^{-3} \frac{f^{-\gamma_{\mathrm{DM}}}}{T} \left( \frac{1400 \ \mathrm{MHz}}{\nu} \right)^{2},
\end{equation}
where $A_{\mathrm{DM}}$ and $\gamma_{\mathrm{DM}}$ are the amplitude and spectral index of the dispersion noise respectively.

\subsection{Samplers}
As described above, the parameter space of even a single pulsar is multidimensional and 
we use techniques of stochastic sampling to infer the noise properties of pulsars. 
To compare among different samplers, we use nested sampling (\cite{skilling2006}) as well as 
MCMC methods, where the latter is also conventionally used in 
inference from pulsar timing data (\cite{justin_ellis_2017_1037579}). 
We have made use of the modular nature of the analysis code \texttt{Enterprise} and 
incorporated different nested samplers to be used with the Likelihood function available within the code. 
We also use the native PTMCMC sampler, both with and without message-passing-interface (MPI) (\cite{mpi40}), making 
a thorough study of performances from different kinds of samplers. 
We briefly describe the individual samplers used in this paper below.
\subsubsection{PTMCMC}
Markov Chain Monte Carlo (MCMC) (\cite{Raftery, gamerman}) is one of the commonest methods to stochastically sample a parameter space. 
Furthermore, Parallel Tempering (\cite{PhysRevLett.57.2607, geyer}) is incorporated to explore the parameter space at different \emph{temperatures}, 
thereby enabling a denser sampling. PTMCMC is natively used in the pulsar timing software \texttt{Enterprise}. MCMC 
directly samples the posterior distribution and after the intial stage, called \emph{burn-in}, gathers samples which are 
the representative posterior samples. In this paper, we have used in addition to PTMCMC, also its MPI-enabled version (henceforth, PTMCMC-MPI) 
and we notice a speedup of around a factor of two in most cases. 
\subsubsection{PyMultiNest}
Conventionally, the nested sampling method samples the prior by distributing \emph{live points} and exploring the parameter space 
by finding higher regions of likelihood. Each live point forms a contour on the likelihood surface 
which gets updated as live points corresponding to lower likelihood values get replaced 
by ones associated with higher likelihood values. Ref.~\cite{multinest} updated this method by forming regions on 
the likelihood surface and associating them to multiple multidimensional ellipsoids. Furthermore, this has been made more user-friendly 
by introducing a \texttt{Python} interface in~\cite{Buchner:2014nha} called PyMultinest. In this paper, we use the parallelised version of the same 
by interfacing it with the MPI protocol. 
\subsubsection{Dynesty}
The nested sampling method described above is known as \emph{Static Nested Sampling}. 
In addition, the Dynesty sampler (\cite{dynesty}) also includes \emph{Dynamic Nested Sampling}. Throughout our paper, we have however 
used the Static sampler from within Dynesty. 
The configuration we have when using Dynesty relies on constructing 
the ellipsoids as implemented in \texttt{MultiNest} and as such differs only in the use of the 
paralleslisation through MPI as we now parallelise sampling the prior. 
In addition, the decision of when to contruct multiple bounds differs in Dynesty as 
opposed to \texttt{MultiNest}. Further information may be found also in~\cite{dynesty-docs}. Our implementation 
follows the call to Dynesty as in~\cite{Smith:2019ucc} and we have adopted 
the approach of parallelisation as in the pubicly available \texttt{pBilby} code.
\subsubsection{UltraNest}
UltraNest (\cite{ultranest}) is a newly introduced nested sampling algorithm. It is designed to ensure accurate sampling of the parameter space, especially in the cases of widely separated minima or tightly correlated parameter density distribution for which multi-ellipsoidal algorithms such as \texttt{MultiNest} have been shown to fail. UltraNest utlises the \texttt{Radfriends} (\cite{radfriends}) algorithm along with flexible penalisation schemes which are dynamically reconfigured to allow resampling of previously sampled regions. We have utilised the Reactive nested sampling algorithm of UltraNest for our test, as we found no discernible benefits from using the static version in our initial testing. It should be noted that the hybrid frequentist and Bayesian approach of standard PTA analysis means this article presents a restrictive comparision for UltraNest as this algorithm is expected to perform better with very large numbers of model parameters.
\begin{figure}
    \centering
    \includegraphics[scale = 0.6]{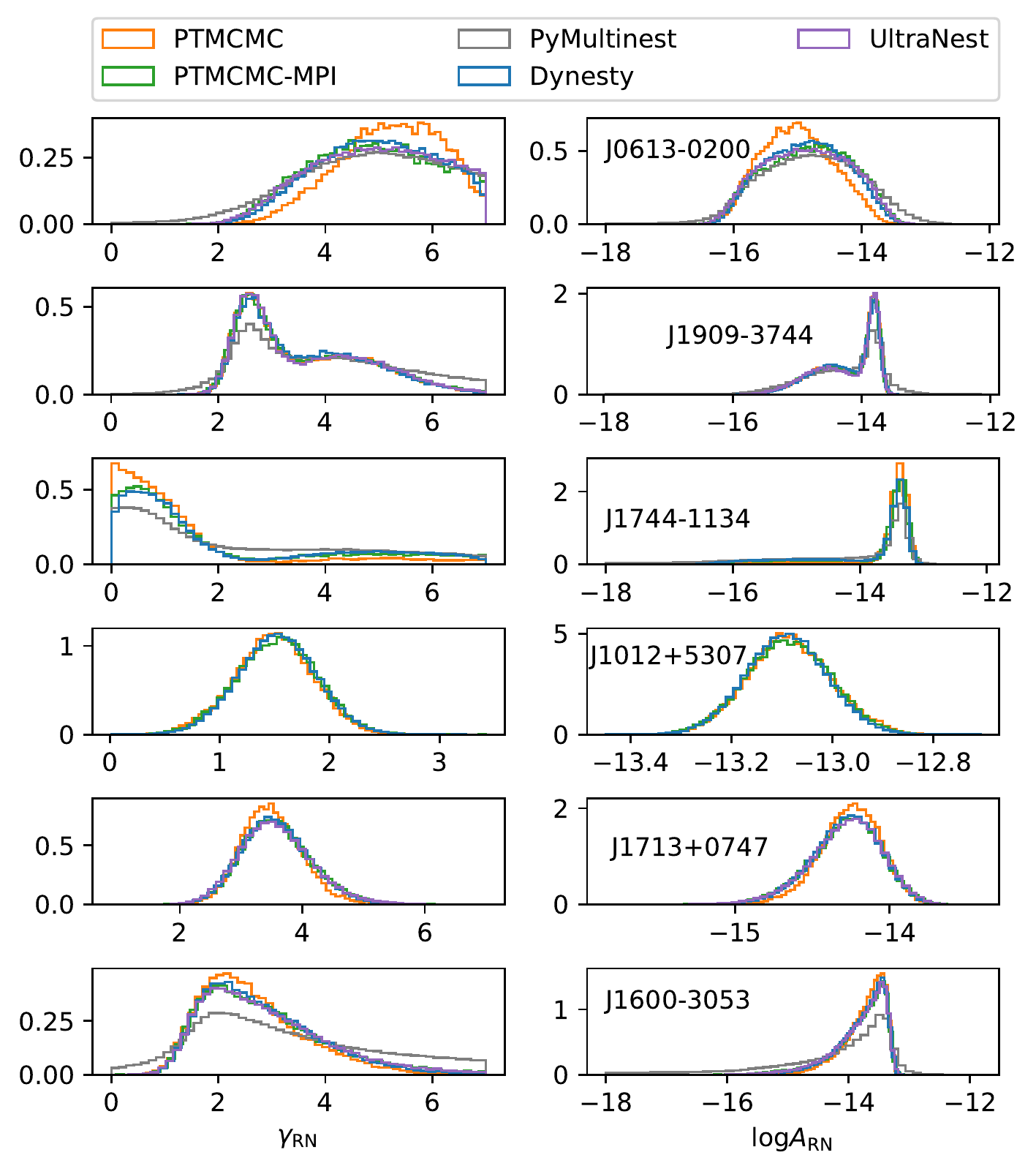}
    \caption{Posterior PDFs across different samplers representing inferred red noise models from SPNA on each of the 6 pulsars from EPTA-DR2. 
             We do not show the results from PyMultiNest on J1713+0747 and J1012+5307 and UltraNest results on 
             J1744-1134 and J1012+5307. These did not finish after several months.}
    \label{fig:spna_rn}
\end{figure}
\begin{figure}
    \centering
    \includegraphics[scale = 0.6]{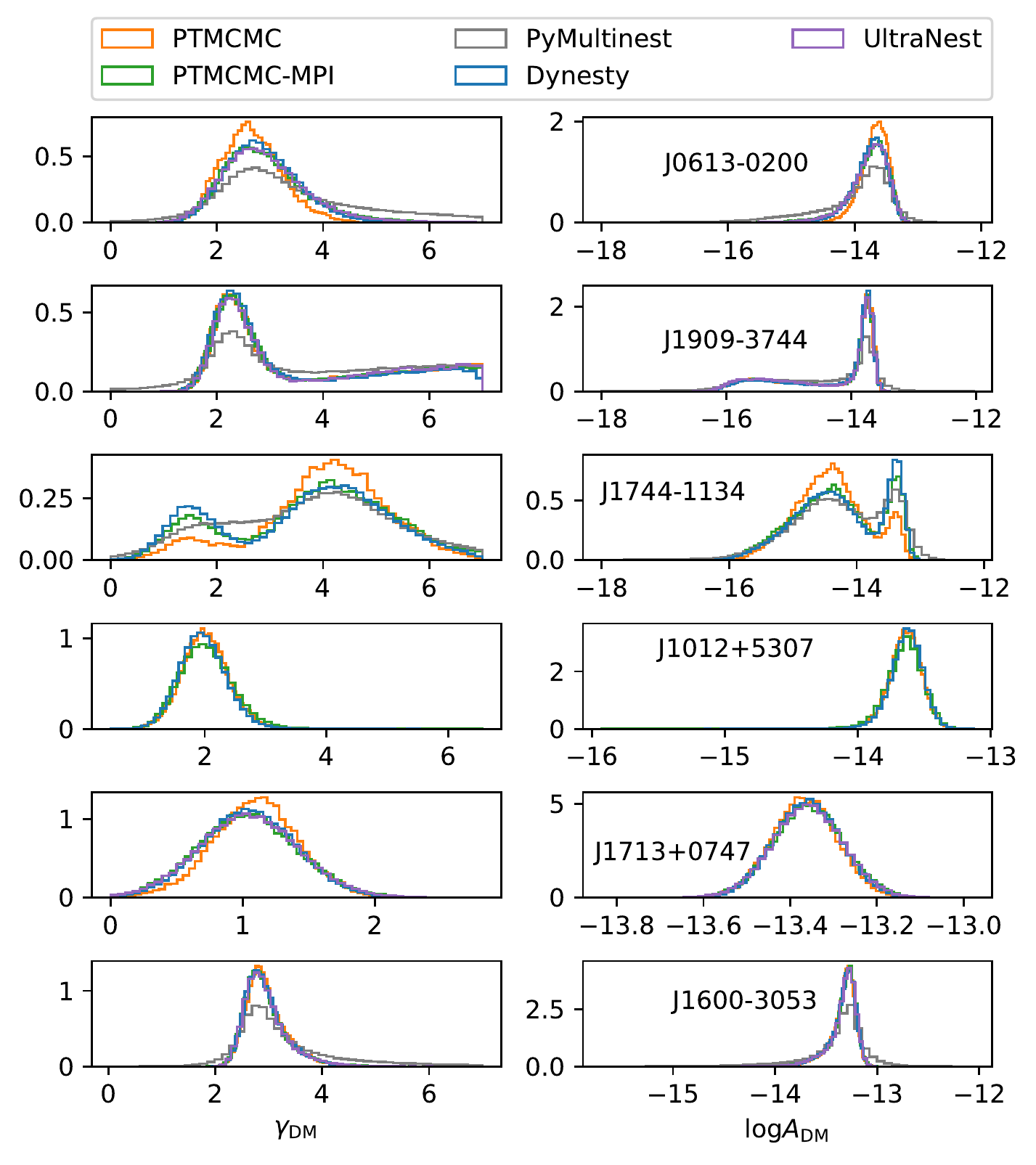}
    \caption{Posterior PDFs across different samplers representing inferred dispersion measure noise models from SPNA on each of the 6 pulsars from EPTA-DR2. 
             We do not show the results from PyMultiNest on J1713+0747 and J1012+5307 and UltraNest results on 
             J1744-1134 and J1012+5307. These did not finish after several months.}
    \label{fig:spna_dm}
\end{figure}
\section{Results}
\label{sec:results}
We present the results from the SPNA analyses on the 6 pulsars from the EPTA and 
present results from the full PTA analyses from the simulated dataset. For the SPNA analyses, 
we have recorded the time taken by each sampler. For the PTA analysis, we focus only on the fastest of 
the nested samplers and use PTMCMC for comparison between two types of samplers. 
The intrinsic parameters being sampled over and their respective prior ranges are given in Tab.~\ref{tab:priors}.
\begin{table}
\centering
\begin{tabular}{c  c}
\hline
  \multicolumn{1}{c}{Parameter} & \multicolumn{1}{c}{Prior range} \\ \hline
 $\log{A_{\mathrm{RN}}}$ & [-18, -10]  \\ 
 $\gamma_{\mathrm{RN}}$ & [0,7] \\
 $\log{A_{\mathrm{DM}}}$ & [-18, -10]  \\ 
 $\gamma_{\mathrm{DM}}$ & [0,7] \\
 $\log{A_{\mathrm{GWB}}}$ & [-18, -10]  \\ 
 $\gamma_{\mathrm{GWB}}$ & [0,7] \\
\hline
   
\end{tabular}
\caption{Prior ranges used; for SPNA runs, the $\mathrm{RN}$ and $\mathrm{DM}$ related parameters are being varied for each pulsar whereas the additional 
         $\mathrm{GWB}$-related parameters are varied in case of the PTA analysis.}
\label{tab:priors}
\end{table}

\subsection{SPNA}
We present results only on RN and DM as these directly affect the TOAs and are 
intrinsic to the pulsars. Fig.~\ref{fig:spna_rn} shows the amplitude and spectral index of the RN noise models 
inferred from the respective 6 EPTA pulsars. The models are presented in the form of posterior Probability Density Functions (PDFs). 
Fig.~\ref{fig:spna_dm} shows the same for the DM noise models. In each case, we show results obtained from different samplers; we show results from PTMCMC, PTMCMC-MPI, PyMultiNest, Dynesty and UltraNest. 
In case of the nested samplers, we have used 4096 live points and have used $2 \times 10^6$ posterior samples for the PTMCMC-based runs for each pulsar. 
Since we have used the samplers each time in combination with the 
generic package \texttt{Enterprise} the likelihood model therefore remains the same and the results show the robustness of the sampling as the PDFs 
are very consistent with each other. We quantify the differences in PDFs by giving the values of \emph{Kolmogorov-Smirnov} (KS) statistic (\cite{Kolmogorov,Smirnov}) in 
Table.~\ref{tab:spna-ks}. If the cumulative distributions corresponding to two posterior distributions $p_1(x)$ and $p_2(x)$ are $P_1(x)$ 
and $P_2(x)$ respectively, the KS statistic is the largest difference:
\begin{equation}
 \mbox{KS} = \mbox{sup}_x | P_1(x) - P_2(x) |.
 \label{eqn:ks}
\end{equation}
From the above definition, the KS value always lies between 0 and 1. If we find the values to be closer to 0, we may consider 
the underlying PDFs $p_1(x)$ and $p_2(x)$ to be very close to each other. 
A source of differences in PDFs is however the stochasticity of the algorithm itself; this is in addition to 
inherent differences between the PDFs being compared. To quantify for this and establish a threshold from the stochasticity itself, for each parameter of each pulsar and for each sampler, we generated 20 sets of resampled posterior samples and computed the KS statistic values between all combinations of these 20 data sets. The maximum KS statistic arising from this study for all parameters and sampler for each pulsar is always $\sim 10^{-3}$, the highest values overall being for J1744-1134, and the $\log{A_{\mathrm{DM}}}$ parameter, $\sim 0.007$. So, from Tab.~\ref{tab:spna-ks}, we take values $> 10^{-3}$ to signify a difference in the PDF arising inherently. 
From Tab.~\ref{tab:spna-ks}, the highest KS value is $\sim$ 0.3 between 
PTMCMC and PyMultiNest for J0613-0200 as well as J1744-1134. We can conclude that from these values, the PDFs among different samplers 
are indeed quite close to each other as is also visually noted from Fig.~\ref{fig:spna_rn} and Fig.~\ref{fig:spna_dm}. 
This leads us to conclude that the parameter estimates obtained from the data are indeed indicative of a physical process and not 
an artifact of sampling.  
We notice PTMCMC combined with MPI gives a speedup in all cases, and while that is a significant gain in runtimes, we note that the algorithm, when coupled with MPI, is different from the native PTMCMC. 
PTMCMC, when used in a single core, does not do parallel tempering (the name is a misnomer in this case). It is only when coupled with MPI, that there is a single temperature per thread and the parallel tempering kicks in\footnote{We thank Michael Keith for pointing this out to us.} 
Moreover, from Figs.~\ref{fig:spna_rn} and~\ref{fig:spna_dm}, we note that PTMCMC-MPI results are closer to those obtained with parallel nested samplers. This is likely because the multiple chains running with different temperatures in case of PTMCMC-MPI,  allow a more exhaustive exploration of the parameter space, making the final posterior PDFs closer to those obtained using nested samplers. 
We also note that, among nested samplers, UltraNest and Dynesty show excellent agreement, whereas PyMultiNest distributions tend to be slightly different. While all the nested samplers that we have used in this work rely on the underlying algorithm \texttt{MultiNest}, there are subtle differences among PyMultiNest, Dynesty and UltraNest. UltraNest and Dynesty have slight improvements over \texttt{MultiNest} (and therefore PyMultiNest) and our results suggest that \texttt{MultiNest} in itself is probably not good enough to sample the complex and high dimensional parameter 
space of the pulsars except in the simplest cases. It may be worth trying to do PyMultiNest analyses with finer settings, however that would not be a one-to-one comparison amongst samplers as is our goal here.
In this analysis, we have only changed the sampler. A robust check of the likelihood function would be to keep the sampler the same 
and change the likelihood definition. The existing software \texttt{TempoNest} (\cite{Lentati:2013rla}) is independent of 
\texttt{Enterprise} and defines the likelihood function independently. It inherently uses the \texttt{MultiNest} sampler, a check of the likelihood 
function may be to repeat an analysis with \texttt{TempoNest} and \texttt{Enterprise} coupled with PyMultiNest. 
This will be studied in a future publication. 
Finally, in Tab.~\ref{tab:spna-times} we note the walltime in hours taken by each sampler on the same data. We also note the number of dimensions, $\mathrm{N_{dim}}$, and number of TOAs, $\mathrm{N_{TOAs}}$ for each pulsar. 
We have used the same machine for all SPNA runs to have a fair comparison of walltimes. A reason for Dynesty's speedup is also 
the parllelisation of the prior-sampling, as mentioned in~\cite{Smith:2019ucc}. 
Specifically, for the pulsars J1713+0747 and J1012+5307, we were unable to get the sampler to 
converge after these runs took at least 80 days and we do not present their results. 
From Tab.~\ref{tab:spna-times}, we note that PyMultiNest becomes unusable for most pulsars; while this is mostly likely due to the increased dimensionality; indeed the missing pulsars for PyMultiNest have $\mathrm{N_{dim}}=56$ and $\mathrm{N_{dim}}=58$, this is likely also a combination of the high $\mathrm{N_{dim}}$ and the large $\mathrm{N_{TOAs}}$. 
In Fig.~\ref{fig:spna_rn} and Fig.~\ref{fig:spna_dm}, all nested samplers and PTMCMC-MPI were run in parallel in one compute node using 47 sockets, where each compute node has the specification of 
\texttt{Intel(R) Xeon(R) Gold 6252N CPU @ 2.30GHz 35.75 MB} with 192GB memory and 48 sockets in total. PTMCMC in itself was run using a single socket in a compute node.
\begin{table*}
\begin{tabular}{c|c|c|c|c|c|c|c|c|c}
\hline
  \multicolumn{2}{p{4cm}|}{} & \multicolumn{2}{p{2cm}|}{PTMCMC vs. PTMCMC-MPI} & \multicolumn{2}{p{2cm}|}{PTMCMC vs. PyMultiNest} & \multicolumn{2}{p{2cm}|}{PTMCMC vs. Dynesty} & \multicolumn{2}{p{2cm}}{PTMCMC vs. UltraNest}\\ \hline
  Pulsar & Parameter & Red Noise & DM Noise & Red Noise & DM Noise  & Red Noise & DM Noise & Red Noise & DM Noise  \\
\hline
 $\mathrm{J0613-0200}$  & $\gamma$ & 0.154 & 0.141 & 0.172 & 0.291 & 0.119 & 0.13 & 0.132 & 0.145 \\
        & $\log{\mathrm{A}}$ & 0.151 & 0.137 & 0.173 & 0.291 & 0.116 & 0.127 & 0.129 & 0.137 \\ \hline
 $\mathrm{J1909-3744}$  & $\gamma$ & 0.01 & 0.009 & 0.094 & 0.117 & 0.023 & 0.041 & 0.012 & 0.016 \\ 
        & $\log{\mathrm{A}}$ & 0.009 & 0.006 & 0.074 & 0.120 & 0.020 & 0.043 & 0.012 & 0.018 \\ \hline
 $\mathrm{J1600-3053}$  & $\gamma$ & 0.079 & 0.038 & 0.196 & 0.185 & 0.062 & 0.035 & 0.086 & 0.037 \\ 
        & $\log{\mathrm{A}}$ & 0.074 & 0.03 & 0.251 & 0.119 & 0.055 & 0.017 & 0.08 & 0.026 \\ \hline
 $\mathrm{J1012+5307}$  & $\gamma$ & 0.042 & 0.042 & - & - & 0.042 & 0.050 & - & - \\
        & $\log{\mathrm{A}}$ & 0.017 & 0.037 & - & - & 0.034 & 0.042 & - & - \\ \hline
 $\mathrm{J1713+0747}$  & $\gamma$ & 0.073 & 0.095 & - & - & 0.062 & 0.077 & 0.072 & 0.088 \\ 
        & $\log{\mathrm{A}}$ & 0.073 & 0.057 & - & - & 0.064 & 0.036 & 0.072 & 0.049 \\ \hline
 $\mathrm{J1744-1134}$  & $\gamma$ & 0.129 & 0.114 & 0.291 & 0.153 & 0.154 & 0.154 & - & - \\ 
        & $\log{\mathrm{A}}$ & 0.113 & 0.112 & 0.295 & 0.166 & 0.136 & 0.150 & - & -  \\ \hline
\end{tabular}
\caption{Results on KS statistics on 6 pulsars from SPNA results from different samplers. The runs which ended up being unfinished after months do not have KS statistics' values associated with 
         them and are given as '-'. The values are shown differently for the red noise and dispersion measure models, and for the two parameters, 
         the amplitude and spectral index separately.}
\label{tab:spna-ks}
\end{table*}

\begin{table*}
\begin{tabular}{c| c| c| c| c| c | c | c | c }
\hline
  \multicolumn{3}{c|}{Pulsar} & \multicolumn{5}{c}{Sampler} \\ \hline
  Name & $\mathrm{N_{dim}}$ & $\mathrm{N_{TOAs}}$ & PTMCMC & PTMCMC-MPI & PyMultiNest & Dynesty & UltraNest  \\
\hline
 $\mathrm{J0613-0200}$ & 50 & 3022 & 9.91  &  8.82 & 745.69 & 6.83  & 42.25 \\ \hline
 $\mathrm{J1909-3744}$ & 18 & 2817 & 13.61 &  5.60 & 3.40   & 1.68  &  2.11 \\ \hline
 $\mathrm{J1600-3053}$ & 30 & 3345 & 19.06 &  7.16 & 16.27  & 3.5   &220 \\ \hline
 $\mathrm{J1012+5307}$ & 56 & 5837 & 28.80 & 13.94 & -      & 11.92 & -      \\ \hline
 $\mathrm{J1713+0747}$ & 58 & 5052 & 29.11 & 14.82 & -      & 15.95 &141.32 \\ \hline
 $\mathrm{J1744-1134}$ & 38 & 1980 & 19.66 &  6.49 & 321    & 4.5   & -      \\ \hline
\end{tabular}
\caption{Walltime in hours for the SPNA runs with each sampler, for each pulsar the number of dimensions and number of TOAs are 
         also given as $\mathrm{N_{dim}}$ and $\mathrm{N_{TOAs}}$ respectively. The unfinished runs' times are shown as `-'. From this table, we note that only 
        PTMCMC and Dynesty are expected to finish within a feasible timescale. Furthermore, when used with MPI, runtimes with PTMCMC can be scaled up. For some pulsars, the speedup obtained is up to a factor of 2.}
\label{tab:spna-times}
\end{table*}
\begin{figure}
    \centering
    \includegraphics[scale=0.5]{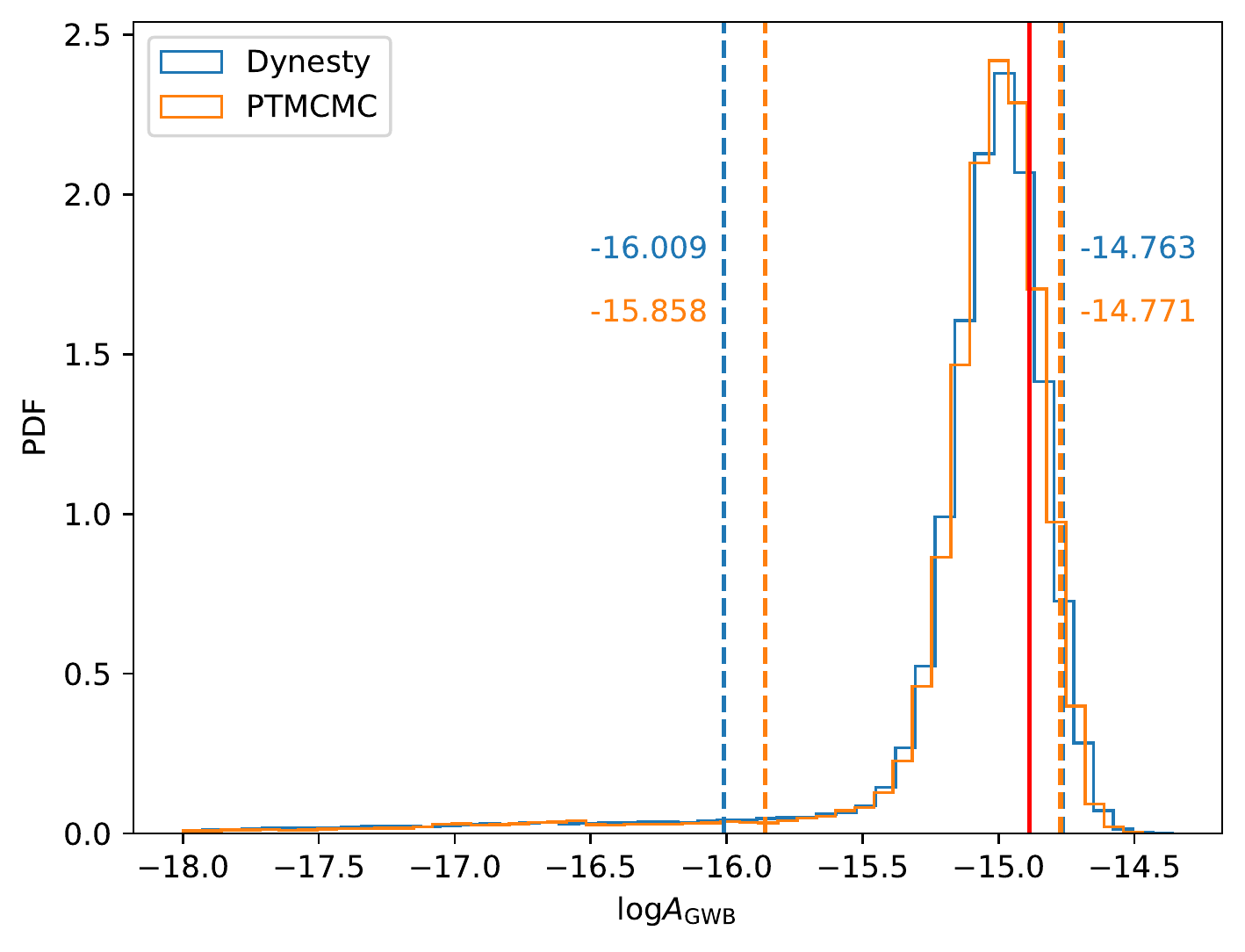}
    \caption{Posterior PDF of the log of GWB amplitude from analysing MDC2 data; the injected value is shown with the red vertical line and is 
             $\log{A_{\mathrm{GWB}}} = -14.89$. 
             The dashed vertical lines refer to the 5\% and 95\% quantiles respectively, we note the injected value always falls within this limit. The spectral index $\gamma$ is kept fixed to 4.33 and for comparison two samplers PTMCMC (orange) and Dynesty (blue) are being run, showing good agreement. For each sampler, the respective values of quantiles (shown in orange for PTMCMC and in blue for Dynesty) also lie very close to each other. 
             }
    \label{fig:pta_gwb}
\end{figure}

\begin{figure}
    \centering
    \includegraphics[scale=0.62]{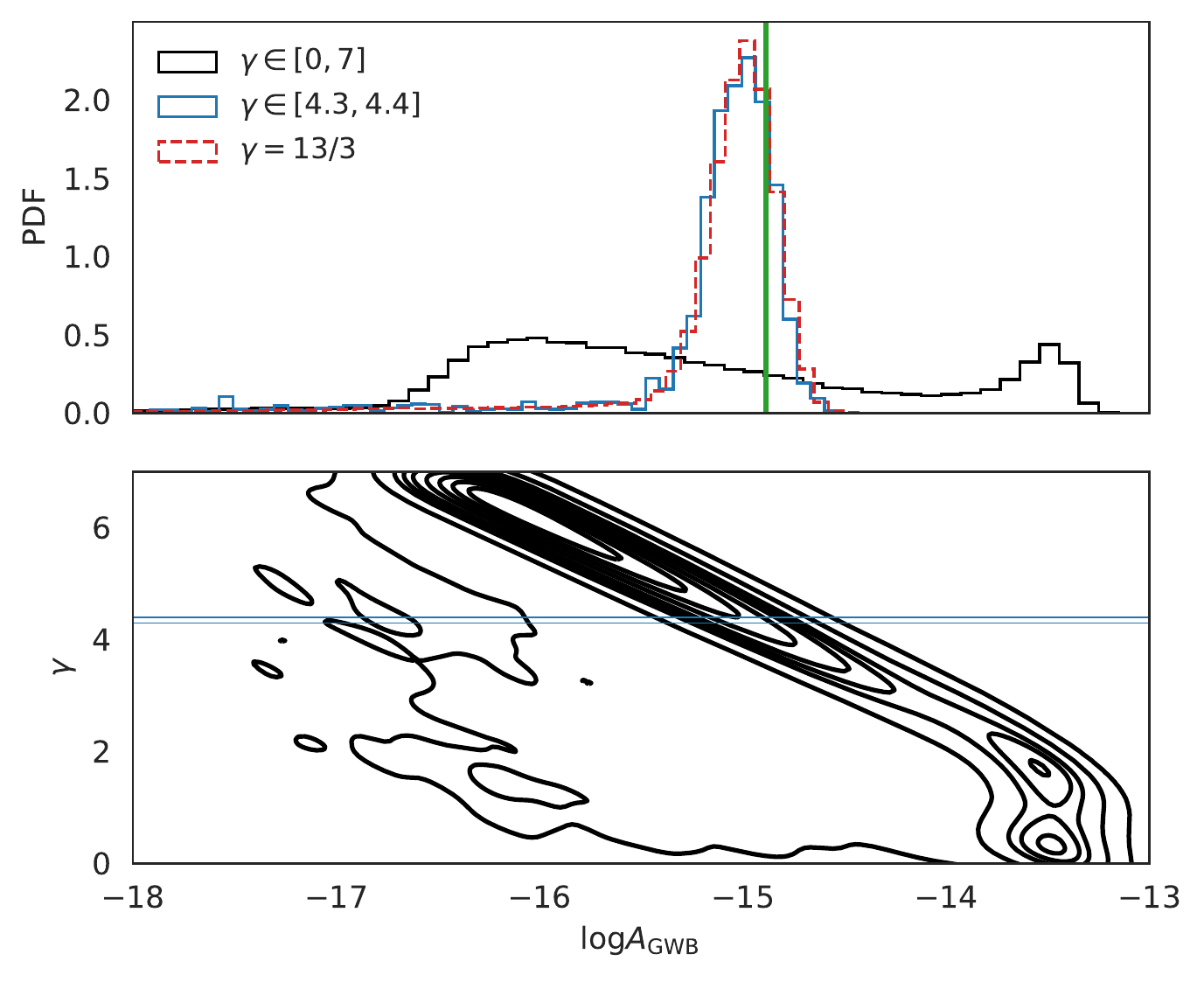}
    \caption{Comparison between varying and fixing the spectral index $\gamma$. The top panel shows the 1D posterior PDFs of the 
             log of the GWB amplitude, $\log{A_{\mathrm{GWB}}}$. The three posteriors correspond to when the spectral index, $\gamma$ is varying in the sampling (black), when the spectral index is fixed to 13/3 in the run (dashed, red) and when the $\log{A_{\mathrm{GWB}}}$ is restricted to the indices of $\gamma$ corresponding to a narrow range of $[4.3, 4.4]$ (blue) in the varying-$\gamma$ run. The injected value of $\log{A_{\mathrm{GWB}}}=-14.89$ is hown as a green vertical line. 
             The bottom panel shows the two-dimensional 
             plot of $\gamma$ versus $\log{A_{\mathrm{GWB}}}$ when $\gamma$ is being varied in the inference run; the $\gamma$ range of $[4.3,4.4]$ corresponding to the blue PDF in the upper panel is shown in blue horizontal lines. 
             We show only the results from Dynesty here as 
             Fig.~\ref{fig:pta_gwb} already shows good agreement between PTMCMC and Dynesty.}
    \label{fig:pta_gwb_comp_var_fix_gam}
\end{figure}

\subsection{PTA}
In this Section, we choose the two fastest samplers from Tab.~\ref{tab:spna-times}. In addition to being the fastest, we also 
use one nested sampler (Dynesty) and one MCMC sampler (PTMCMC) for consistency checks between two different methods of sampling. 
Since we do an analysis 
on all 33 pulsars together which form the pulsar timing array in the IPTA-MDC2, 
we fix the WN parameters to their \texttt{TEMPO2} fit values 
to make the analysis computationally feasible. We analyse simulated data where a GWB has been injected as 
mentioned in Sec.~\ref{sec:data}. 
The injected value of the GWB amplitude is picked up by the resulting PDFs of the GWB amplitude by both samplers 
as shown in Fig.~\ref{fig:pta_gwb}. The figure shows the results when the analysis is done by keeping the 
spectral index of the GWB power spectrum fixed to 4.33. In addition, we repeated the analysis by varying 
both the spectral index and the GWB amplitude.This is shown in Fig.~\ref{fig:pta_gwb_comp_var_fix_gam}. The upper panel shows the results of the GWB amplitude when $\gamma$ is varied 
and kept fixed. In the lower panel the amplitude for the varying $\gamma$ case is plotted by choosing the 
amplitude values corresponding to those of $\gamma$ lying between 4.3 and 4.4 and the resulting PDF looks very similar 
to the case when $\gamma$ is kept fixed to 4.33.

In addition, we perform model selection between a GWB and a CRN, using both samplers. 
With PTMCMC, we use the hypermodel approach, available within \texttt{Enterprise} to extract a Bayes' factor in favour of one of the two 
models. 
The model selection remains inconclusive from the values of Bayes' factors obtained with either sampler. 
Recently Ref.~\cite{Chalumeau:2021fpz} 
also used these two samplers to get model selection results. The values obtained from both samplers 
are given in Tab.~\ref{tab:mod_select}. This shows that we are unable to assign a model to the data even when the data 
is ideally simulated, contains no DM noise and contains a GWB signal of considerable amplitude $A_{\mathrm{GWB}}= 1.3 \times 10^{-15}$. 
This problem of model selection will therefore become even more important in real data which will 
additionally contain unmodelled noise. 
Further, we note the uncertainties in Tab.~\ref{tab:mod_select} and the slightly higher uncertainty values associated with the Dynesty runs. 
While for hypermodel sampling, one run suffices to assign a value of Bayes' factor to a model, with the nested sampling approach, we 
have to resort to separate runs for each model to get a Bayes' factor. The error is therefore added in quadrature and adds up in the case of the runs done with Dynesty. 
In Sec.~\ref{sec:conclusion}, we suggest a method to be able to assign a threshold value of Bayes' factor to claim a detection 
from real data.

\begin{table}
\centering
\begin{tabular}{c| c| c  }
   & $B^{\mathrm{GWB}}_{\mathrm{CRN}}$ & $B^{\mathrm{Fix} \ \gamma}_{\mathrm{Vary} \ \gamma}$  \\
\hline
 Dynesty & 0.534 $\pm$ 1.148  & 0.945 $\pm$ 1.147 \\ \hline
 PTMCMC & 1.279 $\pm$ 0.018  & 0.955 $\pm$ 0.011 \\ \hline
   
\end{tabular}
\caption{Model selection results: Bayes' factors between models GWB and CRN compared with samplers PTMCMC and Dynesty.}
\label{tab:mod_select}
\end{table}
\section{Conclusions}
\label{sec:conclusion}
We have compared different algorithms to sample the parameter space of the pulsar likelihood. We have used samplers to  
infer models from six single pulsars as well as a PTA comprised of thirty three pulsars. In each case, we note good agreement between different 
sampling algorithms and from estimates of run-time as well as to maintain a balance between different ways of sampling, propose the use 
of PTMCMC and Dynesty as preferred methods for future inference from pulsar timing data.

In future we will generate data with randomized sky positions for the pulsars, commonly referred to as sky-scrambling (\cite{Taylor:2016gpq}), to resemble a realistic realisation of measured data and will construct a 
distribution of Bayes' factors to build a `background'. Furthermore, we will inject GWB signals and infer their properties and compare the 
resulting Bayes' factors distribution. This will likely give us an idea of the threshold Bayes' factor to claim a detection of a GWB, if present 
in data. This work is in progress and will be published separately.

\section*{Acknowledgements}
ASa, GS and AS acknowledge financial support provided under the European
  Union's H2020 ERC Consolidator Grant ``Binary Massive Black Hole Astrophysics'' (B Massive, Grant Agreement: 818691). 
  ASa further thanks the Alexander von Humboldt foundation for financial assistance. 
  This work has benefitted from discussions within working groups of EPTA and IPTA; particularly the authors would like to thank 
  Gregory Desvignes, Bhal Chandra Joshi, A. Gopakumar and Boris Goncharov for their comments. 
  JA acknowledges support by the  Stavros Niarchos Foundation (SNF) and the Hellenic Foundation for Research and Innovation (H.F.R.I.) under the 2nd Call of ``Science and Society'' Action Always strive for excellence -- ``Theodoros Papazoglou'' (Project Number: 01431). 
  JWM is a CITA Postdoctoral Fellow: this work was supported by Ontario Research Fund - research Excellence Program (ORF-RE) and the Natural Sciences and Engineering Research Council of Canada (NSERC) [funding reference CRD 523638-18].

\section*{Data Availability}

The paper has made use of data on 6 single pulsars, results from whose analyses have been published by the EPTA in~\cite{Chen:2021rqp} and MDC data simulated by the IPTA, available in \href{https://github.com/ipta/mdc2/}{\color{blue}https://github.com/ipta/mdc2/}.



\bibliographystyle{mnras}
\bibliography{refs} 




\bsp	
\label{lastpage}
\end{document}